\documentclass[nofootinbib,preprint,aps,pra,floatfix]{revtex4-1}
\usepackage[dvips]{graphicx}
\usepackage{amssymb,amsfonts,amsmath}
\usepackage{dcolumn}% align table columns on decimal point
\usepackage{bm}% bold math
\usepackage{textcomp}
\usepackage{multirow}
\usepackage{array}
\usepackage[usenames]{color}

\usepackage[british]{babel}

\begin{document}

\title{Thin layered drawing media probed by THz time-domain spectroscopy}

\author{J. Tasseva$^{1,2}$, A. Taschin$^{1}$, P. Bartolini$^{1}$, J. Striova$^{3}$, R. Fontana$^{3}$, and R. Torre$^{1,4}$}
\email{torre@lens.unifi.it}
\affiliation{
$^1$European lab. for Non-Linear Spectroscopy (LENS), Univ. di Firenze, via N. Carrara 1, I-50019 Sesto Fiorentino, Firenze, Italy.\\
$^2$Institute of Optical Materials and Technolgies “Acad. J. Malinowski”, Bulgarian Academy of Sciences, Acad. G. Bonchev Street, blok 109, 1113 Sofia, Bulgaria\\
$^3$Istituto Nazionale di Ottica, INO-CNR, Largo Fermi 6, I-50125 Firenze, Italy.\\
$^4$Dip. di Fisica e Astronomia, Univ. di Firenze, via Sansone 1, I-50019 Sesto Fiorentino, Firenze, Italy.}

\preprint{version 03}
\date{\today}

\begin{abstract} 
Dry and wet drawing materials were investigated by THz time-domain spectroscopy in transmission mode. Carbon-based and iron-gall inks have been studied, some prepared following ancient recipes and others using current synthetic materials; a commercial ink was studied as well. We measured the THz signals on thin film of liquid inks deposited on polyethylene pellicles, comparing the results with the thick pellets of dried inks blended with polyethylene powder. This study required the implementation of a new experimental method and data analysis procedure able to provide a reliable extraction of the material transmission parameters from a structured sample composed of thin layers, down to thickness of few tens of micrometers. 

THz measurements on thin ink layers enabled the determination of both the absorption and the refractive index in an absolute scale in the 0.1~-~3 THz range, as well as the layer thickness. THz spectroscopic features of a paper sheet dyed by one of the iron-gall inks were also investigated. 
Our results showed that THz time-domain spectroscopy enables to discriminate the various inks on different supports, including the application on paper, together with a proper determination of the absorption coefficients and indices of refraction. 
\end{abstract}

\maketitle
\section{Introduction}

THz pulse imaging and spectroscopy is an emerging non-invasive method for the characterization of cultural heritage artefacts that provides complementary information to traditional analytical tools. With respect to X-rays or UV radiation, THz radiation presents lower risks in terms of molecular stability due to the non-ionizing property as well as the ability of inducing very low thermal stress. Transmission/reflection geometries, time- or frequency domain set-ups are constantly developed for a variety of applications ranging from pharmaceutical chemical mapping, safe security to atmospheric sensing. The potential to provide non-destructive information in the cultural heritage filed has been demonstrated on objects such as paintings~\citep{Seco_13}, manuscripts~\citep{Fukunaga_08,Abraham_09,Labaune_10}, mural paintings~\citep{Walker_13}, and metal alloys~\citep{Jackson_15} and stone~\citep{Krugener_15}. In particular, Bardon et al~\citep{Bardon_13} have reported the investigation of black inks by THz spectroscopy. 

T-ray technologies are supported by a series of commercial off-the-shelf systems that enable THz spectroscopic investigations. Designed for few specific applications other than cultural heritage, these systems are not open to a full control of signal detection and processing. Therefore, a customized set-up was built providing for a more flexible and powerful application. 

In this work, we explored artworks drawing materials with THz-Time Domain Spectroscopy (THz-TDS) extending the investigation to thin layers of inks, that was never realized previously. We developed a new specific experimental method and data analysis to disentangle the multiple reflection signals.
Thanks to a high signal to noise ratio and the  accurate analysis implemented, our measurements enable the calculation of the absolute absorption coefficient and index of refraction of the materials, as well as the sample thickness down to tens of microns both in single layer and bilayer configurations.

\section{Materials and Experimental procedures}

The inks studied, both iron gall and carbon black inks (Table~\ref{table}) are commonly used in artworks and can be prepared either following ancient recipes (Giovanni Alcherio 1411) or from synthetic materials \cite{merrifield_67}. 
They have been sampled from pellets or thin layers; dried inks were blended with polyethylene (PE) powder with a concentration of approximately $33~wt.\%$ and pressed to form pellets of $13.2~mm$ diameter and a thickness of about $1~mm$, wet inks were deposited on $10~\mu m$ thick PE pellicles to form films with thickness of the order of tens of micrometers. Moreover, the iron gall ink (recipe B) was also studied when applied to pure cotton paper (Zecchi, Firenze).
We investigated the THz spectrum of other inks (i.e. red coachineal, red ochre, indigotin-based and white lead) as well as the individual chemical constituents (i.e. Iron (II) sulfate and gallic acid); these results are not reported here and will be discussed in a future publication.

\begin{table}[htb]
  \begin{center}
    \caption{Compositions and recipes for the preparation of the studied inks. The rough materials have been purchased from Bizzari S.A.S. Firenze, Italy.}
    \label{table}
   
    \begin{tabular}{|>{\centering\arraybackslash}m{1cm}|>{\centering\arraybackslash}m{0.3 cm}|m{13cm}|}
 \hline
 \multirow{5}{*} {Black}
	  
& 1 & \textbf{Iron gall ink with Arabic gum} (recipe A): 70 mL water, 10~ mL white wine, 10 mL red vinegar, 5 g powdered oak galls, 1.25 g Arabic gum, 1.25 g FeSO$_4\cdot$7H$_2$O. \\ \cline{2-3}
	  	  	      
& 2 & \textbf{Iron gall ink without Arabic gum} (recipe A): 70 mL water, 10~mL white wine, 10 mL red vinegar, 5 g powdered oak galls, 1.25 g FeSO$_4\cdot$7H$_2$O. \\ \cline{2-3}
	 	          
& 3 & \textbf{Iron gall ink with Arabic gum} (recipe B): 14 mL distilled water, 1.14 g gallic acid, 0.29 g Arabic gum, 0.29 g FeSO$_4\cdot$7H$_2$O. \\ \cline{2-3}
 			 
& 4 & \textbf{Iron gall ink} commercial black ink (Zecchi, Firenze)\\ \cline{2-3}
 	 
& 5 & \textbf{Carbon-based ink}: 14 mL red wine, 0.57 g carbon black (vegetal), 0.71 g Arabic gum\\
     \hline
    \end{tabular}

  \end{center}
\end{table}

\begin{figure}[htb]
\centering
\includegraphics[width=0.8\textwidth]{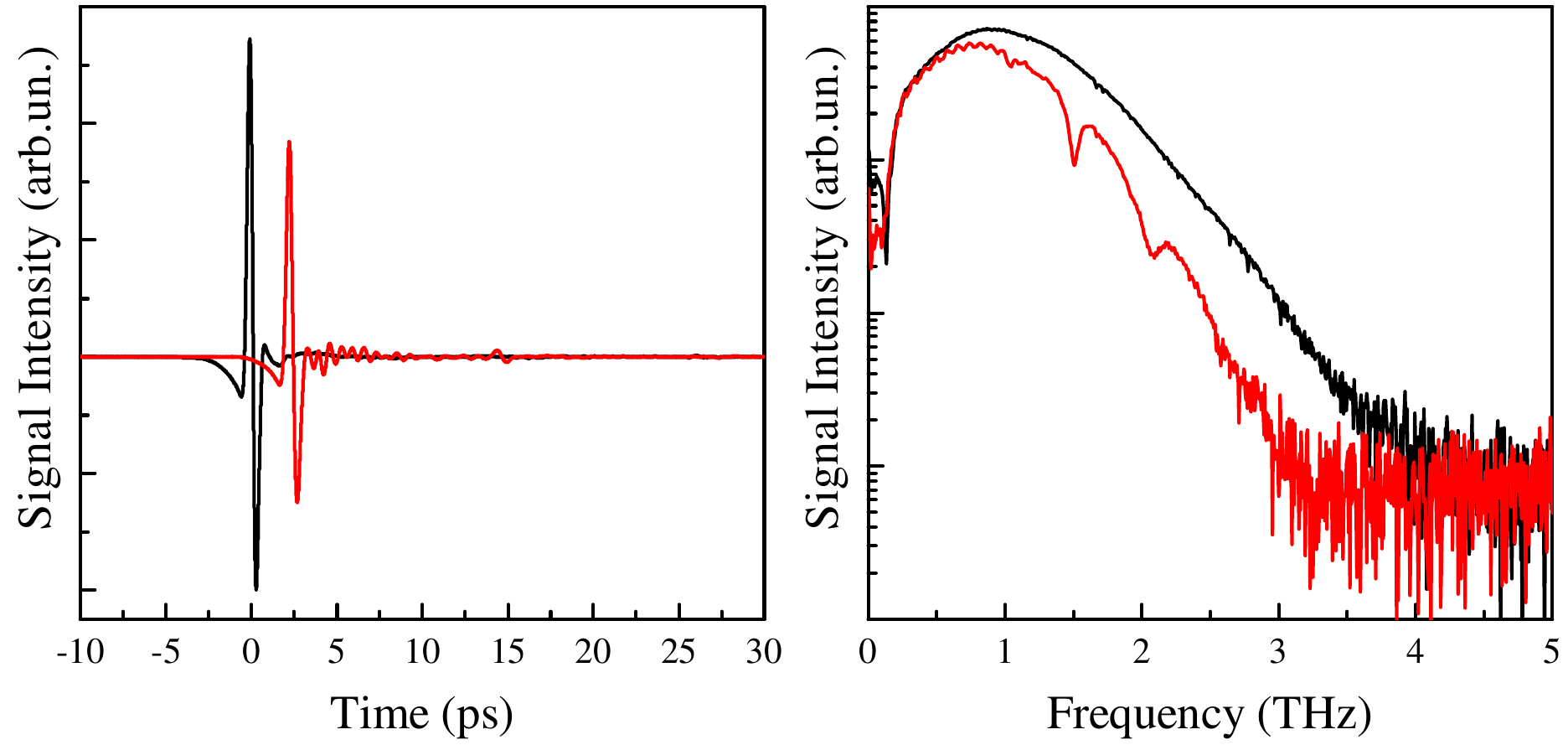}
\caption{Left graph: typical time evolution of the THz electric field of the reference and ink signal from a pellet sample (Black 3 ink), the second pulse at around 15 $ps$ is due to internal reflections of the THz pulse between the pellet surfaces; Right graph: amplitude spectra obtained by Fourier transform of the signals on the left. The regular oscillations clearly visible in the lower frequency range of the spectrum are due to the internal reflections. This effect must be carefully considered in order to extract the real spectrum of the sample parameters.}
\label{data-example}
\end{figure}

Our table top THz-TDS system enables measurements in 0.1-4 THz range in transmission configuration. The sample and reference THz signals are cyclically acquired; for every sample scan we took a reference scan. Each single scan is a $300$ second long acquisition with a continuous motion of the probe delay line at a velocity of $0.5~mm/s$. Each couple of sample and reference signals is Fourier transformed and their ratio, averaged over all the data, gives the experimental transfer functions defined in the next paragraph. A typical couple of reference and sample signals with their amplitude spectra are shown in figure~\ref{data-example} where  the Black 3 ink signal from the pellet sample is reported.
A detailed description of the experimental set-up and the sample preparation is reported in the Supplementary Information.

\section{THz transmission and sample parameters}

The spectroscopic properties in the THz range of a material can be obtained by measuring the sample and the reference signals.  
The ratio between the Fourier transforms of the THz field transmitted through the sample, $E_t \left( \omega\right)$, and of the incident field, $E_{i}\left( \omega\right)$, refereed to as the \textit{Transfer Function} of the material, $H\left( \omega\right)$, expresses how a plane wave of frequency $\omega$ is modified by the absorption and refraction from the medium it encounters. For a homogeneous dielectric slab of thickness $d$ and complex refractive index $\hat{n}_s$, surrounded by nitrogen, and for normal incidence, the theoretical expression of the transfer function can be written as~\citep{Withaya_14}:
\begin{align}
\label{Hfun1}
	H(\omega) = \frac{E_t(\omega)}{E_i(\omega)} =  \tau\tau' \exp{\left\lbrace  -i\left[ \hat{n}_s(\omega)-n_0\right]  \frac{\omega d}{c}\right\rbrace } \cdot FP(\omega)
\end{align}
\begin{align}
\label{HfunFP}
	FP(\omega) =\sum\limits_{m=0}^\infty \left\lbrace  \rho'^2 \exp{\left[-2i\hat{n}_s(\omega) \frac{\omega d}{c}\right]} \right\rbrace^m
\end{align}
%\left\lbrace  1-\rho'^2 \exp{\left[-2i\hat{n}_s(\omega) \frac{\omega d}{c}\right]} \right\rbrace^{-1}
where 
\begin{equation}
\tau  = 2n_0/(n_0+\hat{n}_s)
\end{equation}
is the nitrogen-sample complex transmission coefficient  and 
\begin{equation}
\tau'=2\hat{n}_s/(n_0+\hat{n}_s)~~,~~\rho' = (n_0-\hat{n}_s)/(n_0+\hat{n}_s)
\end{equation}
are the sample-nitrogen complex transmission and reflection coefficients for normal incidence; with $\hat{n}_s=n_s(\omega)-ik_s(\omega)$, where $n_s(\omega)$ is the refractive index, $k_s(\omega)$ the extinction coefficient, and $n_0$ the real refractive index of nitrogen. 
Yet in eq.s~\ref{Hfun1} and \ref{HfunFP}, $c$ is the vacuum speed of light and $FP(\omega)$ represents the Fabry-P\'{e}rot effect due to the multiple reflections inside the sample.

In a TDS transmission experiment the material can be characterized by measuring the experimental transfer function, $H_{exp}( \omega)$, given by the ratio between the complex Fourier transform of the sample signal and the reference signal, from which using eq.s ~\ref{Hfun1} and \ref{HfunFP} the refractive index, $n_s\left( \omega\right)$, the absorption coefficient, $\alpha_s(\omega)=2\omega k_s(\omega)/c$ and the thickness could be in principle extracted. 

However, a complete analytical solution of the eq.s~\ref{Hfun1} and \ref{HfunFP} is not available, so the extraction of the sample parameters must be pursued following alternative methods. These are strongly dependent on the nature of the THz signal measured.

If the $FP$ pulse reflections are distinguishable and separable from other contributions in the sample temporal signal, the transfer function can be obtained from the THz signals cutting off the pulse reflections. This simple method enables an immediate extraction of material parameters, but the sample thickness must be known a priori and the detection time-window is limited by the cut (i.e. the $n_s(\omega)$ and $k_s(\omega)$ spectra are characterized by a lower resolution).

 If the $FP$ pulse reflections are close in time and partially superimposed with the main pulse, due to a short optical path, the cutting process can't be applied. Also in the case where the reflection peaks are well separated but the sample signal shows a long time evolution after the main peak, because of a structured absorption of the medium, see figure~\ref{data-example}, the simple method can give wrong evaluations of the material parameters. In these cases, the solution methods are based on an iterative process of calculation~\citep{Withaya_05,Pupeza_07,Scheller_09,Scheller_09b,Scheller_11,krimi_16}. 

We implemented a new method of data analysis, described in details in the Supplementary Information, which is an iterative fitting process based on a polynomial fit of the transmission parameters that enables the extraction of the real physical material parameters (i.e. index of refraction, absorption coefficient and layer thickness). This method applies to both the free standing single slab or layer (i.e. our pellet samples) and a bilayer system, (i.e. the layered inks on PE pellicles).

\section{Results and discussion}

\subsection{Ink-PE Pellets}
\begin{figure}[htb]
\centering
\includegraphics[width=0.9\textwidth]{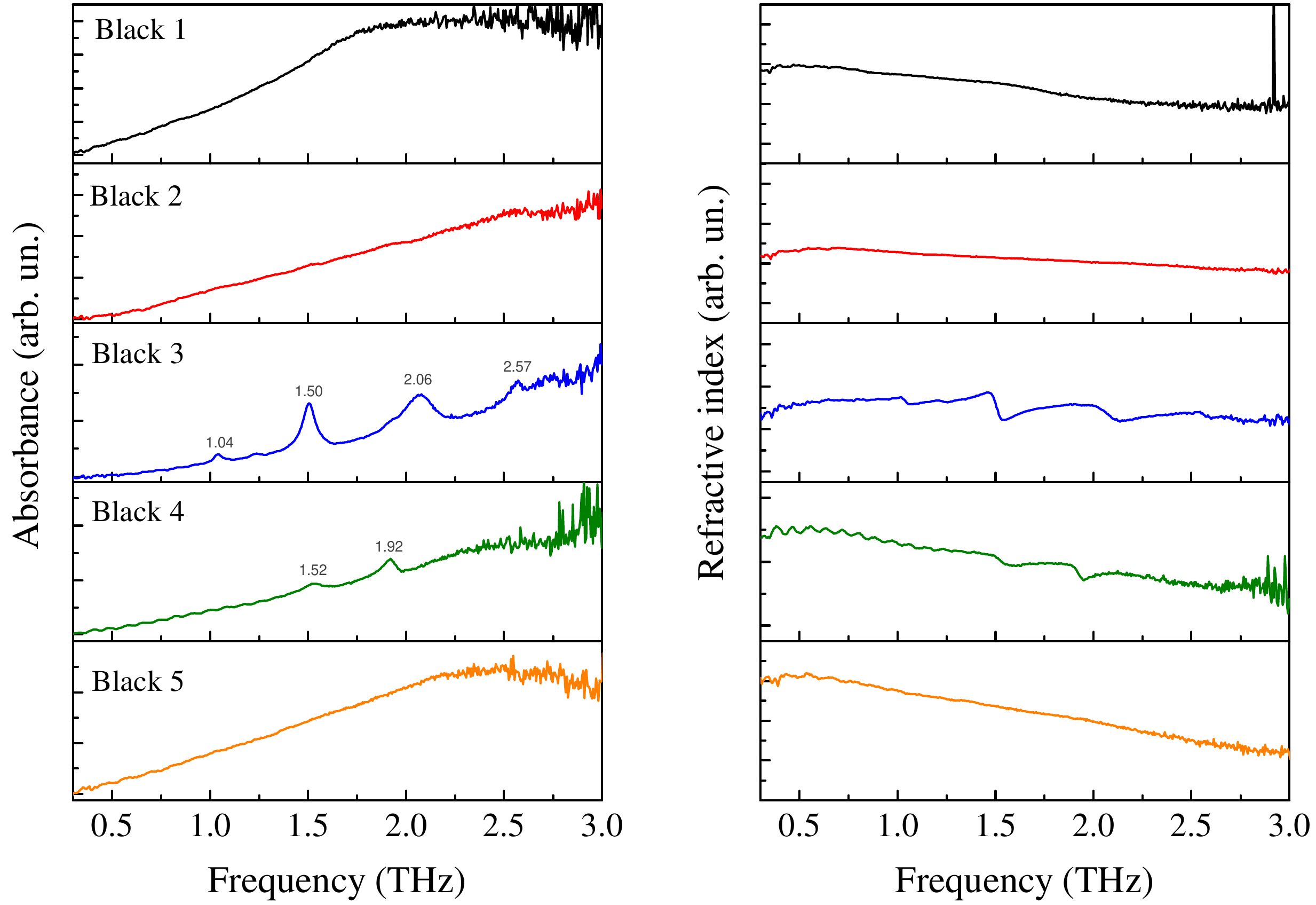}
\caption{THz spectra for the iron gall inks measured on pellets. The Left Panel shows the absorption coefficients; the pigments prepared following the ancient recipe (Recipe A) containing oak galls (Black 1 and Black 2) and the carbon-based one (Black 5) do not shown any particular spectral feature contrary to what observed in the inks containing synthesized gallic acid (Black 3 and Black 4) which instead exhibit well distinguished peaks. The Right Panel reports the frequency behaviour of the refractive indices measured on the same series of black ink pellets. The dispersive features of the refractive index for the Black 3 and Black 4 inks confirm what measured in the respective absorbance spectra.}
\label{black-ink-pellets}
\end{figure}

Figure~\ref{black-ink-pellets} shows the absorbance spectra and frequency evolution of the refractive indexes for all five black inks, see Table \ref{table}, measured on pellets. The inks prepared following the old recipe of mixing mainly iron(II) sulfate and oak galls powder (i.e. recipe A used for Black 1 and 2) show a featureless response in the THz range. The comparison between these two spectra suggests that the Arabic gum weakly contributes to the spectral response with a smooth increase of the absorption in the higher frequency range. On the other hand, ink containing synthetic gallic acid (i.e. Black 3 prepared according to recipe B) and commercial ink (i.e. Black 4), show structured spectra with several peaks. Carbon black ink, Black 5, does not exhibit any spectral features. Our results are in agreement with other THz spectroscopic investigations on similar drawing media~\citep{Bardon_13}. 
The inks prepared following recipe A, Black 1 and 2, do not show any features related either to gallic acid or to iron(II) sulfate. Actually, it is known that under hydrolysis gallic acid is extracted from the oak galls and the related spectroscopic features should appear in the THz spectra. The absence of spectroscopic signatures of gallic acid and iron sulfate is probably due to the chemical processes taking place in the ink preparation. More likely, the iron-gallic complexes oxidized supporting the formation of iron-tannic complexes that are expected not to show any spectroscopic features in the 0.15 - 3 THz range~\citep{Bardon_13}. 
In the spectra of Black 3, prepared according to recipe B that uses the ferrous sulfate and synthesized gallic acid instead of oak-galls powder, we can clearly see the spectroscopic signature of the gallic acid (peaks at about 1.04, 1.50, 2.06, and 2.57 THz) while no ferrous sulfate contribution is present in the spectra. This is likely due to the low molar ratio between the two chemical compounds, about 0.1, which prevents the observation of iron sulfate features~\citep{Bardon_13}. 
Differently, the commercially available iron gall ink, Black 4, shows a couple of features typical of the iron sulfate (peaks at about 1.52, 1.92 THz) and lacks those of gallic acid. 

\subsection{Ink Layers on PE}

The pellet samples provide very good spectra of the drawing media, as shown in the previous paragraph, nevertheless they have several drawbacks. First of all, these spectra are not acquired on drawing materials in form of inscriptions, brushstrokes, ecc.

Moreover, due to difficulties in calculating the molar concentration of inks in pellets, no absolute measurements of the molar absorption coefficient and refractive index may be obtained. Furthermore, scattering processes due to inhomogeneities in the sample must be taken into account as, for example, the band distortions in the absorption spectra, named Christiansen effect, due to the phenomena of Mie scattering~\citep{Franz_08}.
The direct investigation of ink layers by THz-TDS overcomes the previous problems and enables the study of drawing media in their common forms. Nevertheless, this turns out to be a quite difficult task especially if the absolute measurement of the THz transmission parameters is searched for. 

\begin{figure}[htb]
\centering
\includegraphics[width=1\textwidth]{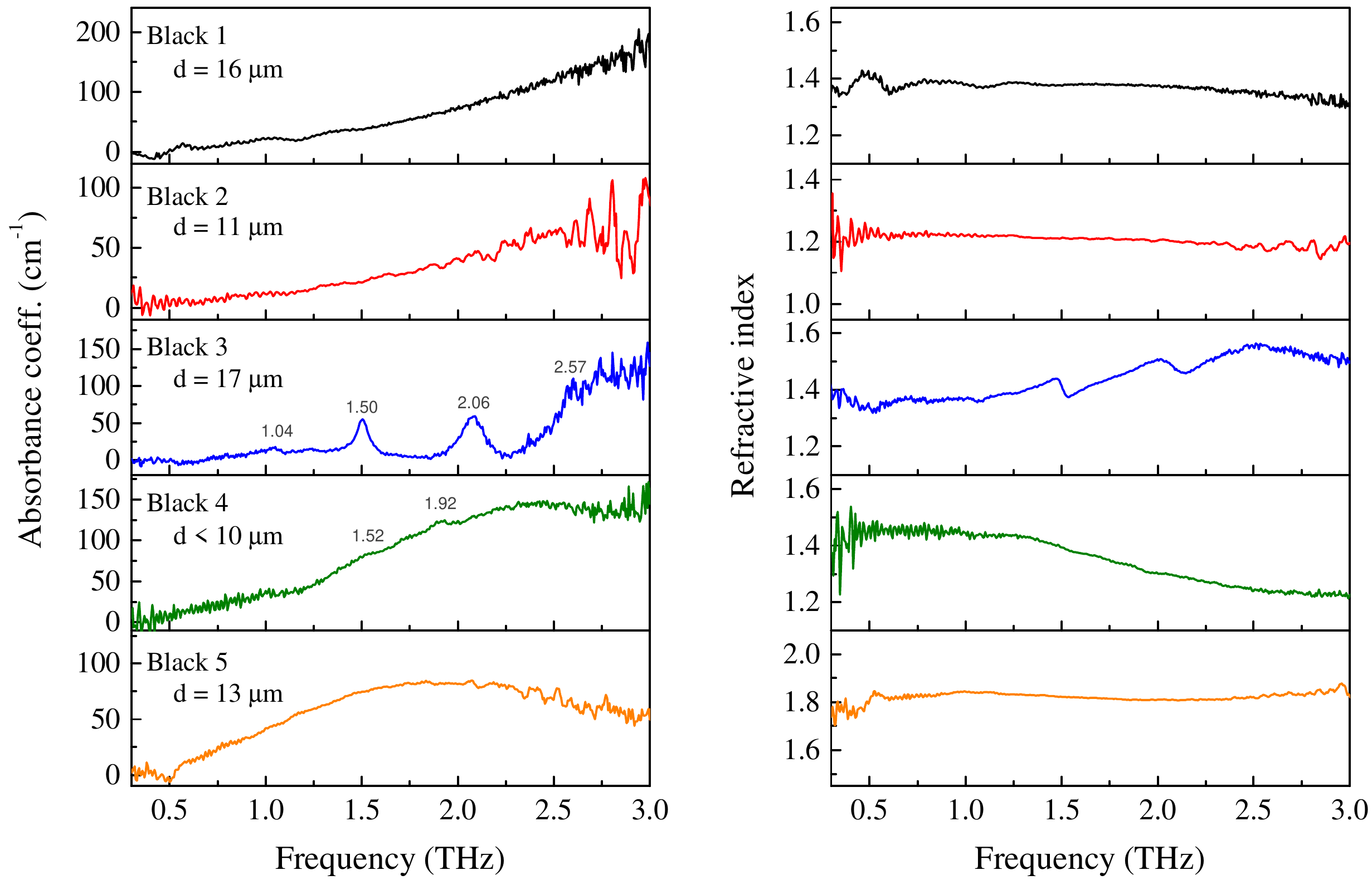}
\caption{Absorption coefficient (left panel) and refractive index (right panel) as a function of frequency for the series of black inks on PE pellicles. Thickness, d, measured from each sample, is also reported. The present THz-TDS investigations enable to obtain absolute values of both THz transmission parameters and of the drawing media thickness.}
\label{black-ink-films}
\end{figure}

In figure~\ref{black-ink-films} we show the optical parameters\footnote{To conform us with the previous literature, the transmission characteristics of the material are referred to as ``optical parameters'' even if they concern the THz spectral region that is outside the commonly defined optical region} in the THz range of thin bi-layer samples made by deposition of thin ink films on PE pellicles, as obtained from the data analysis of the TDS investigation. The analysed inks are from the same series reported in the Table \ref{table}.
Our experimental apparatus and data analysis procedure give spectra of very good signal-to-noise ratio even for thin film samples. We highlight that the determined optical parameters and the layer thickness are in absolute scale. 

Ink layer data confirm the spectral features shown by the pellet samples; the layered inks prepared with oak galls powder (Black 1 and 2) again do not manifest spectral features; the ink prepared by recipe B (Black 3) reveals the presence of gallic acid by peaks in absorbance at 1.50 and 2.06 THz; the commercial iron-gall ink (Black 4) shows the 1.92 THz feature of ferrous sulfate. Therefore the THz-TDS technique proves to be able to distinguish among the various black inks even when they are in the form of thin film.
\subsection{Ink on paper}
\begin{figure}[htb]
\centering
\includegraphics[width=0.9\textwidth]{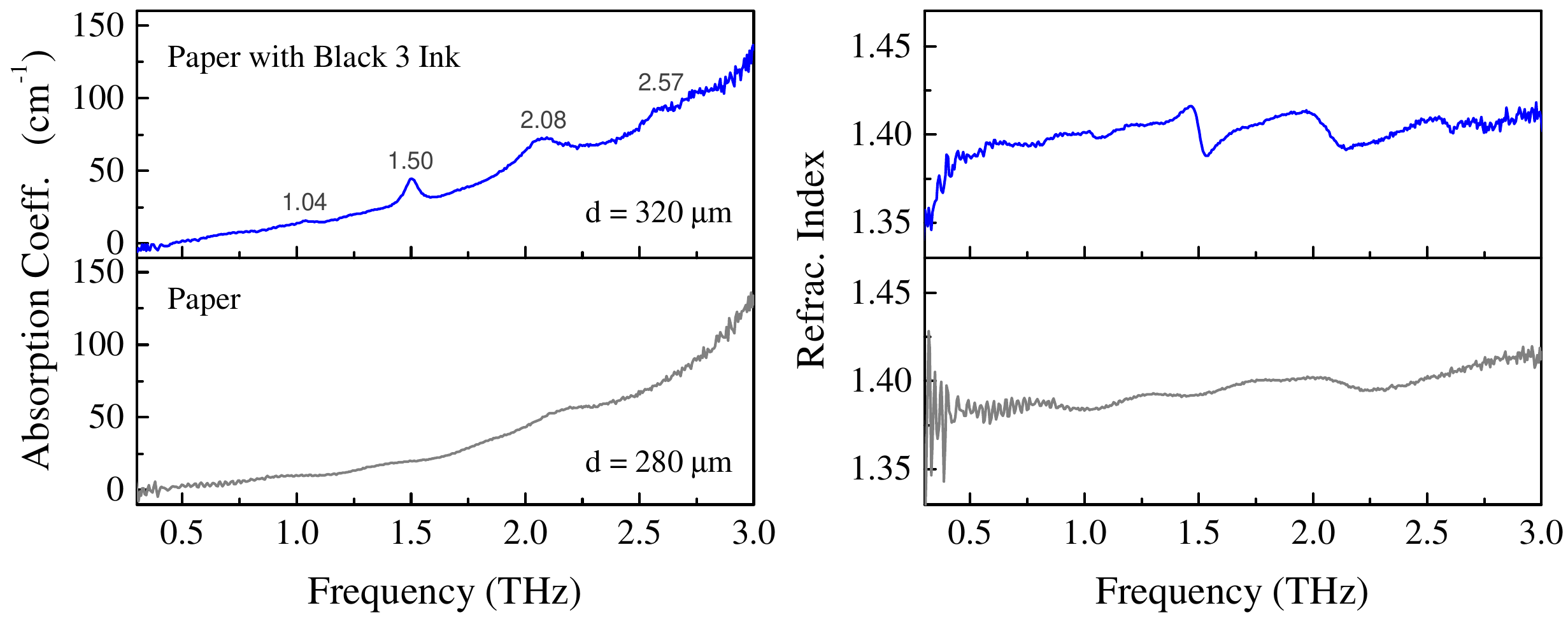}
\caption{Absorption spectra and refraction index of a paper sheet dyed with Black 3 ink. The typical THz spectroscopic signature of Black 3 ink is still clearly visible even when it impregnates its original support.}
\label{black-ink-paper}
\end{figure}

We applied the investigation by THz-TDS experiments on a sample prepared by depositing a few drops of Black 3 ink on a piece of drawing paper. This sample is quite different from the previous ones; in fact the ink penetrates into the paper forming a complex structure where ink particles and paper fibres are interpenetrated. We treated the sample as a homogeneous system. In figure \ref{black-ink-paper} we report the results of data analysis. The paper sheet shows a relatively smooth THz spectrum characterized by some very weak absorption bands, after ink deposition, the spectrum shows clearly the signature of Black 3 ink with a calculated sample thickness of about $40$ $\mu m$.

\section{Conclusions}
We develop an original experimental procedure and a comprehensive method of data analysis to measure the optical parameters of drawing media in the THz spectral range. The THz time-domain spectroscopy  was applied to investigate black inks, as iron-gall and carbon based inks commonly used in artworks. We determined the absorption coefficient and refractive index spectra of pellet and layered ink samples. The THz spectroscopic measurements on thin films of ink enable the determination of the optical parameters in an absolute scale, making the present research unique and pushing the investigation to a complete quantitative analysis of the real-practice problems and suggesting potential applications of THz spectroscopy in the heritage field. We found the THz-TDS techniques able to discriminate between iron-gall black inks prepared either with old recipe or with modern/commercial preparations both in the pellet sample and in layered film. Measurements on inked paper, the most commonly utilized support for inks, confirm that it is possible to discern the spectroscopic features of ink also when it penetrates a paper sheet. The reported results point out some common challenges present in THz spectroscopy and imaging of artworks and suggest the possible methods to overcome them.

\section{Acknowledgement}
We thank very much Dr. C. Riminesi for his support on this research. This work was founded by Regione Toscana, prog. POR-CROFSE-UNIFI-26 and by Ente Cassa di Risparmio Firenze, prog. 2015-0857. Authors of INO-CNR acknowledge IPERION CH project GA 654028, funded by EU community’s H2020-research infrastructure program. We acknowledge M. De Pas, A. Montori, and M. Giuntini for providing their continuous assistance in the electronic set-ups; and R. Ballerini and A. Hajeb for the mechanical realizations.

%\bibliography{Papers_THz}

\begin{thebibliography}{10}

\bibitem{Seco_13}
C.~Seco-Martorell, V.~L\'{o}pez-Dom\'{i}nguez, G.~Arauz-Garofalo,
  A.~Redo-Sanchez, J.~Palacios, and J.~Tejada.
\newblock Goya's artwork imaging with terahertz waves.
\newblock {\em Opt. Express}, 21(15):17800--17805, Jul 2013.

\bibitem{Fukunaga_08}
Kaori Fukunaga, Yuichi Ogawa, Shin'ichiro Hayashi, and Iwao Hosako.
\newblock Application of terahertz spectroscopy for character recognition in a
  medieval manuscript.
\newblock {\em IEICE Electronics Express}, 5(7):223--228, 2008.

\bibitem{Abraham_09}
E.~Abraham, A.~Younus, A.~El Fatimy, J.C. Delagnes, E.~Nguéma, and P.~Mounaix.
\newblock Broadband terahertz imaging of documents written with lead pencils.
\newblock {\em Optics Communications}, 282(15):3104 -- 3107, 2009.

\bibitem{Labaune_10}
J.~Labaune, J.B. Jackson, S.~Pag\'{e}-Camagna, I.N. Duling, M.~Menu, and G.A.
  Mourou.
\newblock Papyrus imaging with terahertz time domain spectroscopy.
\newblock {\em Applied Physics A}, 100(3):607--612, 2010.

\bibitem{Walker_13}
Gillian~C. Walker, John~W. Bowen, Wendy Matthews, Soumali Roychowdhury, Julien
  Labaune, Gerard Mourou, Michel Menu, Ian Hodder, and J.~Bianca Jackson.
\newblock Sub-surface terahertz imaging through uneven surfaces: visualizing
  neolithic wall paintings in \c{C}atalh\"{o}y\"{u}k.
\newblock {\em Opt. Express}, 21(7):8126--8134, Apr 2013.

\bibitem{Jackson_15}
J.~Bianca Jackson, Julien Labaune, Rozenn Bailleul-Lesuer, Laura D'Alessandro,
  Alison Whyte, John~W. Bowen, Michel Menu, and Gerard Mourou.
\newblock Terahertz pulse imaging in archaeology.
\newblock {\em Frontiers of Optoelectronics}, 8(1):81--92, 2015.

\bibitem{Krugener_15}
S.~F. Busch A. Soltani E. Castro-Camus M.~Koch K.~Krügener, M.~Schwerdtfeger
  and W.~Viöl.
\newblock Terahertz meets sculptural and architectural art: Evaluation and
  conservation of stone objects with t-ray technology.
\newblock {\em Scientific Reports}, 5(14842), 2015.

\bibitem{Bardon_13}
Tiphaine Bardon, Robert~K. May, Philip~F. Taday, and Matija Strlic.
\newblock Systematic study of terahertz time-domain spectra of historically
  informed black inks.
\newblock {\em Analyst}, 138:4859--4869, 2013.

\bibitem{merrifield_67}
M.P. Merrifield.
\newblock {\em Original treatises on the arts of painting}.
\newblock Original Treatises on the Arts of Painting. Dover Publications, 1967.

\bibitem{Withaya_14}
Withawat Withayachumnankul and Mira Naftaly.
\newblock Fundamentals of measurement in terahertz time-domain spectroscopy.
\newblock {\em Journal of Infrared, Millimeter, and Terahertz Waves},
  35(8):610--637, 2014.

\bibitem{Withaya_05}
W.~Withayachumnankul, B.~Ferguson, T.~Rainsford, S.~P. Mickan, and D.~Abbott.
\newblock Material parameter extraction for terahertz time-domain spectroscopy
  using fixed-point iteration.
\newblock {\em Proc. SPIE}, 5840:221--231, 2005.

\bibitem{Pupeza_07}
Ioachim Pupeza, Rafal Wilk, and Martin Koch.
\newblock Highly accurate optical material parameter determination with thz
  time-domain spectroscopy.
\newblock {\em Opt. Express}, 15(7):4335--4350, Apr 2007.

\bibitem{Scheller_09}
Maik Scheller, Christian Jansen, and Martin Koch.
\newblock Analyzing sub-100-$\mu$m samples with transmission terahertz time
  domain spectroscopy.
\newblock {\em Optics Communications}, 282(7):1304 -- 1306, 2009.

\bibitem{Scheller_09b}
Maik Scheller and Martin Koch.
\newblock Fast and accurate thickness determination of unknown materials using
  terahertz time domain spectroscopy.
\newblock {\em Journal of Infrared, Millimeter, and Terahertz Waves},
  30(7):762--769, 2009.

\bibitem{Scheller_11}
Maik Scheller.
\newblock Real-time terahertz material characterization by numerical
  three-dimensional optimization.
\newblock {\em Opt. Express}, 19(11):10647--10655, May 2011.

\bibitem{krimi_16}
Soufiene Krimi, Jens Klier, Joachim Jonuscheit, Georg von Freymann, Ralph
  Urbansky, and René Beigang.
\newblock Highly accurate thickness measurement of multi-layered automotive
  paints using terahertz technology.
\newblock {\em Applied Physics Letters}, 109(2):021105, 2016.

\bibitem{Franz_08}
Morten Franz, Bernd~M. Fischer, and Markus Walther.
\newblock The christiansen effect in terahertz time-domain spectra of
  coarse-grained powders.
\newblock {\em Applied Physics Letters}, 92(2):021107, 2008.

\bibitem{Lee_09}
Yun-Shik Lee.
\newblock {\em Principles of terahertz science and technology}.
\newblock Springer, 2009.

\bibitem{MacFarlane_94}
Duncan~L. MacFarlane and Eric~M. Dowling.
\newblock Z-domain techniques in the analysis of fabry--perot \'{e}talons and
  multilayer structures.
\newblock {\em J. Opt. Soc. Am. A}, 11(1):236--245, Jan 1994.

\bibitem{Jin_14}
BiaoBing Jin, CaiHong Zhang, XiaoFang Shen, JinLong Ma, Jian Chen, ShengCai
  Shi, and PeiHeng Wu.
\newblock Extraction of material parameters of a bi-layer structure using
  terahertz time-domain spectroscopy.
\newblock {\em Science China Information Sciences}, 57(8):1--10, 2014.

\end{thebibliography}

\newpage

\appendix

\section*{Supplementary Information}

\subsection{Ink preparation and recipes}

The studied inks were either prepared according to old recipes or purchased from Zecchi (Florence, IT). In Table~\ref{table}, present in the article, we report the compositions of the studied inks and the recipes followed for their preparations.
The black inks are iron gall inks, which were prepared in lab following two main recipes containing either oak galls as a source of the gallo-tannic acid (Recipe A in Table ~\ref{table}) or preliminarily synthesized gallic acid instead (Recipe B). The recipe A is a historic recipe from Giovanni Alcherio 1411  (on the advice of experts in paper conservation from Opificio delle Pietre Dure Institute in Florence) \cite{merrifield_67}; the ingredients (water, wine and vinegar) were mixed together with powdered oak galls and left for $1$ month to allow for the gallo-tannic acid extraction. The mixture was then heated until the volume was reduced by $1/4$. The powdered Arabic gum was added and briefly heated, iron(II) sulfate was added at the end.

\subsection{Samples preparation}
Inks were dried in laboratory conditions, blended with polyethylene (PE) powder (Merck), ground, and pressed under a manual hydraulic press at ~0.8 GPa to form pellets of $13.2~mm$ diameter and thickness of about $1~mm$. The analytic concentration in the PE pellet was set to be approximately $33~wt.\%$ that, for most of the studied inks, has revealed to be an optimal concentration, making detectable any eventual features in a relatively wide spectral range.
PE was found to be an ideal support for absorption spectroscopy in the THz region thanks to its negligible absorption coefficient (below $1~cm^{-1}$ see~\citep{Lee_09} and references therein). In order to study the material's optical properties, wet inks were also deposited on $10~\mu m$ thick PE pellicles (IR sample cards, Sigma-Aldrich) and let dry to form films with thickness of tens of $\mu m$. Moreover, the iron gall ink (recipe B) was also studied when applied on paper. 

\subsection{THz time-domain spectroscopy set-up}
\begin{figure}[htb]
\centering
\includegraphics[width=0.9\textwidth]{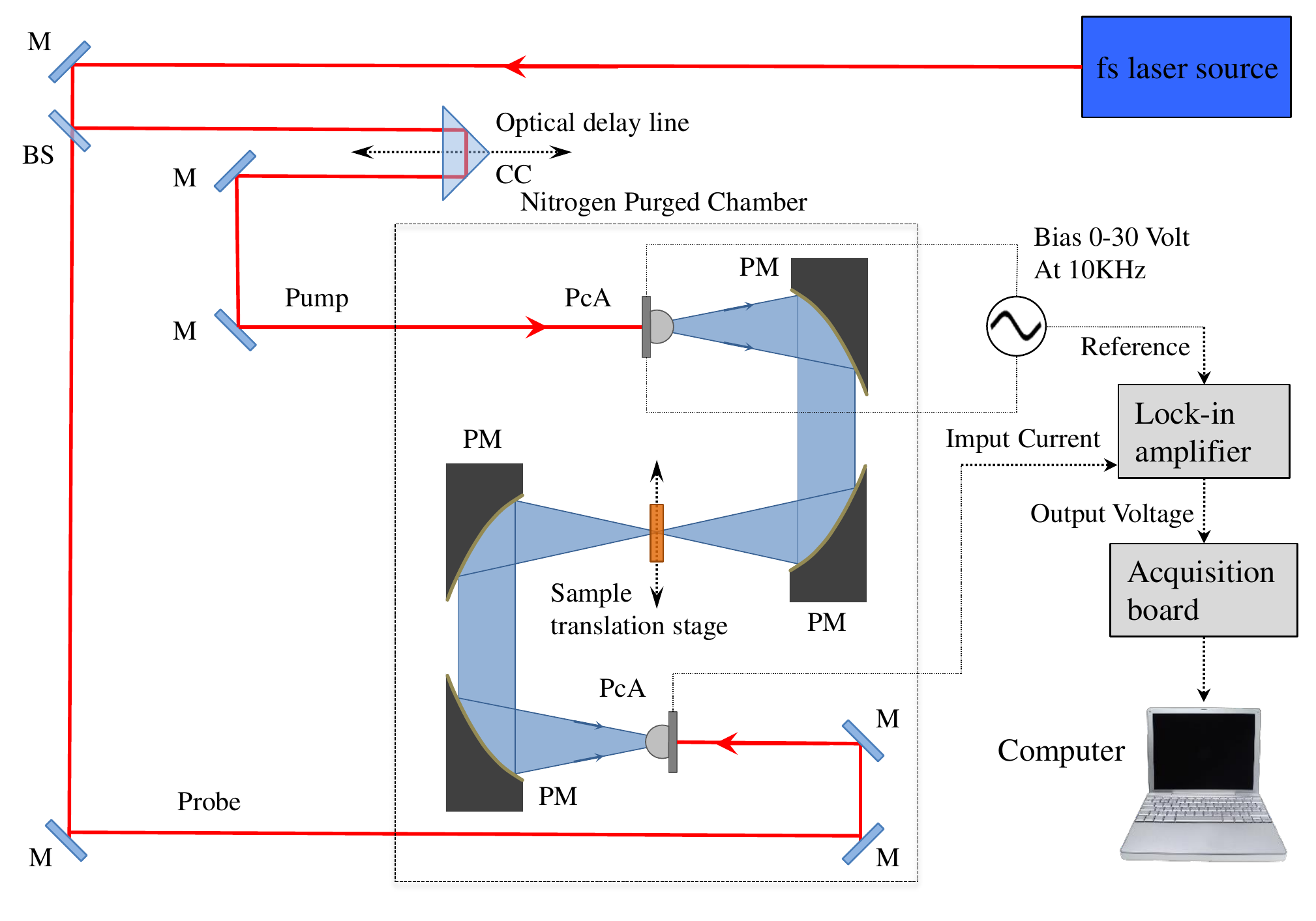}
\caption{
Optical set-up for THz time-domain spectroscopy in transmission configuration: M – mirror, BS – beam splitter, CC – corner cube, PcA – photoconductive antenna, PM – parabolic mirror.}
\label{setup}
\end{figure}

Measurements in 0.1 - 4 THz range were performed with a home-made THz-TDS system in transmission configuration. Figure~\ref{setup} depicts a simplified scheme of our THz-TDS set-up. Optical laser pulses, at $\lambda=780~nm$, with a pulse duration of less than $120~fs$ and repetition rate of $100~MHz$ (produced by a T-light 780 nm fiber laser from MenloSystems), excites a Low-temperature GaAs photoconductive antenna (PcA)~\citep{Lee_09}, which is biased with sinusoidal voltage of $0-30~Volt$ at a frequency of $10~KHz$. The photoexcited carriers are accelerated and shortly after recombine. The abruptly varying photocurrent generates short bursts of electromagnetic radiation with a broad spectrum in the THz region. The emitted THz field is extracted and collected by a hemispherical silicon lens to obtain a divergent beam, which is then collimated and subsequently focused on the sample by a couple of parabolic off-axis mirrors (PMs). The signal transmitted through the sample is again collimated and focused on the detector PcA by a second couple of PMs and an other hemispherical silicon lens optimizes the coupling between the THz filed and the dipole of the antenna. A second optical pulse, the probe, generates photo-excited carriers that are accelerated by the THz field, which acts as bias in the detection PcA. The temporal evolution of the photocurrent amplitude in the detection antenna, obtained by changing the time delay between the pump and probe pulses, is directly connected to the electric field amplitude of the THz radiation. This current is amplified by a lock-in amplifier, locked at the bias frequency of the source antenna, and digitalized by an acquisition board. A home-made software acquires the processed signal together with the reading of the delay line encoder and retraces the final time dependent THz field. The working chamber, containing the whole THz set-up, was purged with nitrogen to eliminate the numerous contributions of water vapour, present at the THz frequencies spanned by the experiment. 
 In order to improve the data quality and reduce the effects of external perturbations during the acquisition (e.g. temperature fluctuations) measurements with and without sample are cyclically repeated. The sample is, therefore, mounted on a motorized translation stage to be moved in front of the beam, and for every sample scan a reference scan is also acquired. Each single scan is obtained by a $300$ second acquisition at a rate of $10~KHz$ with a continuous motion of the probe delay line at the velocity of $0.5~mm/sec$. Each couple of sample and reference signals are Fourier transformed and their ratio is averaged over all the data, thus giving the experimental transfer functions defined in the article.

\subsection{Material parameters extraction from experimental data}

In our work we followed the numerical optimisation algorithm proposed by Scheller et al.~\citep{Scheller_11}. 

The first step is to obtain a rough estimation of $n_s$ and $\alpha_s$. By neglecting the $FP$ term and the imaginary part of the refractive index in the Fresnel coefficients of eq.~\ref{Hfun1}, we get analytical expressions for the optical parameters~\citep{Withaya_14}:
%%\text{arg}
\begin{align}
	n_s(\omega)&=n_0-\frac{c}{\omega d}\text{arg}\left[H(\omega)\right]   \label{ns}\\
	k_s(\omega)&=\frac{c}{\omega d}\left\lbrace ln \left[ \frac{4n_0n_s}{\vert H(\omega) \vert(n_0+n_s)^2}\right] \right\rbrace \label{alphas}
\end{align}
where $\text{arg}\left[H(\omega)\right]$ is the argument of the complex transfer function.
Substituting in $H$ the experimental value $H_{exp}$ and an assumed initial value of $d$ measured with a micrometric screw, we obtain approximated frequency-dependent values of $n_s$ and $\alpha_s$, which are, moreover, affected by fake oscillations due to the neglected $FP$ effect (see fig.~\ref{data-example} in the article). In order to remove these oscillations, Scheller et al.~\citep{Scheller_11} apply to these raw values of $n_s$ and $\alpha_s$ a band stop filter centred at the $FP$ frequency. We instead implement a polynomial fit of the optical parameters, varying the polynomial order and fitting range as long as the real physical frequency behaviour is revealed and the $FP$ oscillations are removed.

After this first estimation of $n_s$, $\alpha_s$, and $d$, we calculate the full theoretical expression of $H(\omega)$, eq.s~\ref{Hfun1} and \ref{HfunFP} in the article, with the summation of the $FP$ limited to the number of reflections appearing in the time window of the measurement, then we compare it with the experimental one to infer new best values for $n_s$, $\alpha_s$, and $d$. Thus the second step is to minimize the function
\begin{equation}\label{deltaH}
\bigtriangleup H=\sum\limits_{\omega}\vert H(\omega)-H_{exp}(\omega)\vert
\end{equation}
with a numerical optimization on the $n_s$, $\alpha_s$ for different fixed values of $d$. We use a Nelder-Mean simplex algorithm with the two scalars $\xi$ and $\psi$:
\begin{align}
n_{s,new}(\omega)&=\xi \left[ n_{s,old} (\omega)-1 \right]+1,\label{parn}\\
k_{s,new}(\omega)&=\psi k_{s,old}(\omega),\label{park}
\end{align}
For every value of $d$, new values of $n_s(\omega,d)$ and $\alpha_s(\omega,d)$ are calculated by eq.s~\ref{ns} and \ref{alphas}, filtered, and then optimized minimizing $\bigtriangleup H$. Plotting the minima of $\bigtriangleup H$ as a function of $d$ we obtain a curve with a minimum in a $d_{min}$ value, which corresponds to the real thickness of the sample. The fitting process is then repeated, starting from the triad, $n_s(\omega,d_{min})$, $\alpha_s(\omega,d_{min})$, and $d_{min}$, but now with the additional parametrization of $d$,  $d=\zeta d$, in order to refine its value.
It is worth to stress that the parametrizations of $n_s$ and $\alpha_s$ through the scalar $\xi$ and $\psi$ do not change their frequency behaviours which are still those inferred from the first step and can be affected by the filtering process.
Thus, as third and final step, as already reported by Scheller et all.~\citep{Scheller_11}, we perform an optimization of the optical parameters at every frequency step $\omega_i$ using the function
\begin{equation}\label{deltaHwi}
\bigtriangleup H(\omega_i)=\vert H(\omega_i)-H_{exp}(\omega_i)\vert
\end{equation}
The starting values for $n_s$ and $\alpha_s$ are the optimal ones found in the previous step, the parametrizations are the same of eq.s~\ref{parn} and \ref{park} with the same algorithm, whilst $d$ is always kept fixed to the optimal value estimated before.
This last optimization reshapes the frequency features of the optical constants that might have been distorted or erased by the first step evaluation and filtering process. 
After the polynomial filtering process, this new set of curves of $n_s$ and $\alpha_s$ can be used again as input values of step two of the optimization cycle. Especially for a sample with a short optical path, the optimization must be repeated several times to find the reliable values of the thickness and the optical constants. 
All the calculations and minimization routines written above by which we analyse all the data reported in this work were performed by executing an in-house developed Matlab code.
What described so far concerns the employed analysis for a free standing single slab or layer, thus relevant for our pellet samples.
In the case of a bilayer system, as in the layered inks on PE pellicles, the optimization process is similar but starts from a different set of equations. 
The analysis can be carried out if the optical properties of at least one of the layers and its thickness are known. In our case the PE pellicle has been firstly characterized as a free standing layer by means of the above analysis: we found $d_{PE}=10~\mu m$, $n=1.40$ frequency independent, $\alpha_{PE}\simeq0$ in the whole probed frequency range.
The first step is to consider the bilayer system as a single layer and obtain effective optical parameters using the approximated eq.s~\ref{ns} and \ref{alphas} with $d=d_1+d_2$, where $d_1$ and $d_2$ are the thicknesses of the two layers. $n_{eff}$ and $\alpha_{eff}$ can be related to the optical constants of the two layers by simple considerations on the refractive index and absorption coefficient. Let us consider the ink layer as 1 and the PE pellicle as the layer 2, we then get:
\begin{align}
n_1&=\frac{1}{d_1}\left[(n_{eff}-n0)(d_1+d_2)-d_2(n2-n0)\right]+n0,\\
k_1&=\frac{1}{d_1}k_{eff}(d_1+d_2)-k_2 \frac{d_2}{d_1},
\end{align}
these have to be filtered from the $FP$ oscillations following the same procedure described above. Then we can calculate the final values of the parameters following the same procedure of optimization and minimization as in the single slab case simply using the correct expression of the transfer function for a bilayer system, which, for waves at normal incidence, can be written as~\citep{MacFarlane_94}:
\begin{align}\label{Hfun2}
H(\omega) &= \frac{E_t(\omega)}{E_i(\omega)} =  \nonumber \\ 
&= \frac
{\tau_{01}\tau_{12}\tau_{20}~e^{ -i\frac{\omega}{c}\left[  d_1\hat{n}_1+d_2\hat{n}_2-n_0\left( d_1+d_2\right) \right] } }
{\left[1-\rho_{21}\rho_{20}~e^{-i\frac{2\omega}{c}d_2\hat{n}_2}\right]
\left[ 1-\rho_{12}\rho_{10}~e^{-i\frac{2\omega}{c}d_1\hat{n}_1}-\frac{\rho_{20}\rho_{10}\tau_{21}\tau_{12}~e^{-i\frac{2\omega}{c}\left( d_1\hat{n}_1+d_2\hat{n}_2\right)}}{1-\rho_{21}\rho_{20}~e^{-i\frac{2\omega}{c}d_2\hat{n}_2}} \right]} 
\end{align}
where $\hat{n}_i$ are the complex refractive indices, $\tau_{ij}$ and $\rho_{ij}$ are the complex transmission and reflection coefficients with $i,j=0$ for nitrogen, being $1$ for ink, and $2$ for PE. This expression includes a $FP$ effect with an infinite number of reflexes from the three interfaces. The experimental transfer function, however, is obtained with temporal signal measured in a finite temporal range, so a number of reflexes restricted to this time window should be considered. Taking into account a finite number of reflections is much more complicated and time-consuming. Anyway, considering that the data temporal range is very long compared to the time delays of the reflections in a sample of two very thin layers and that the intensity of each subsequent reflection decays exponentially in time, we expect that the experimental transfer function can be accurately described by eq.~\ref{Hfun2}~\citep{Jin_14}.

\end{document}